\documentstyle [12pt] {article}
\LARGE
\Large
\huge
\begin{document}
\begin{center}
{ \bf

  ABSTRACTS FOR THE FIELDS INSTITUTE FOR RESEARCH IN MATHEMATICAL SCIENCES
     AND NATO ADVANCED RESEARCH WORKSHOP PROGRAM }

\vspace{7ex}
{\bf        PATTERN FORMATION and
            LATTICE-GAS AUTOMATA   }\\

\vspace{7ex}
	   {\bf JUNE 07-12, 1993 }

Supported by \vspace{1ex}

{\bf The Ontario Ministry of Education and Training and\\
The Natural Sciences and Engineering Research Council of Canada\\
and\\
NATO Advanced Research Workshop Program}

\vspace{7ex}

\leftskip=3true cm{\bf	       ORGANIZING COMMITTEE}\vspace{2ex}

\leftskip=3true cm
{ Anna T. Lawniczak, Guelph University, Canada   }\\

\leftskip=3true cm{ Raymond Kapral, University of Toronto, Canada}\\
\vspace{2true cm}

\leftskip=3true cm{\bf	      SCIENTIFIC COMMITTEE}   \vspace{2ex}

\leftskip=3true cm
{  J.P. Boon, Universit\'{e} Libre Bruxelles, Belgium\\
             G.D. Doolen, Los Alamos National Laboratory, USA \\
             R. Kapral, University of Toronto, Canada	       \\
             A.T. Lawniczak, University of Guelph, Canada	 \\
             D. Rothman, Massachusetts Institute of Technology, USA }
\end{center}

\newpage
{\bf  INVITED SPEAKERS}

\vspace{2.0true cm}

 1. \ {\bf C. Appert}, Universit\'{e} Pierre et Marie Curie, France

 2. \ {\bf B. Boghosian}, Thinking Machines, USA

 3. \  {\bf J.P. Boon}, Universit\'{e} Libre Bruxelles, Belgium

 4. \ {\bf S. Chen}, Los Alamos National Laboratory, USA

 5. \ {\bf E.G.D. Cohen}, Rockefeller University, USA

 6. \ {\bf D. Dab}, Universit\a'{e} Libre Bruxelles, Belgium

 7. \ {\bf A. De Masi}, Universit\a`{a} degli
Studi di L'Aquila, Italy

 8. \ {\bf G.D. Doolen}, Los Alamos National Laboratory, USA

 9. \ {\bf M. Ernst}, University of Utrecht, Netherlands

10. \ {\bf E.G. Flekkoy}, University of Oslo, Norway

11. \  {\bf B. Hasslacher}, Los Alamos National Laboratory, USA

12. \ {\bf F. Hayot}, Ohio State University, USA

13. \ {\bf M. Henon}, CNRS Observatoire de Nice, France

14. \ {\bf D. d'Humi\a`{e}res}, Ecole Normale Sup\a'{e}rieure, France

15.  \ {\bf  N. Margolus}, Massachusetts Institute of Technology, USA

16. \ {\bf R. Monaco}, University of Genova, Italy

17. \ {\bf A. Noullez}, Princeton University, USA

18. \ {\bf S. Ponce--Dawson}, Los Alamos National Laboratory, USA

19. \ {\bf E. Presutti},
 II Universit\a`{a} degli Studi di Roma, Italy

20. \ {\bf R. Rechtman}, Universidad Nacional
Autonoma de Mexico, Mexico

21. \ {\bf D. Rothman}, Ecole Normale Sup\a'{e}rieure, France

22. \ {\bf S. Succi},
 IBM Center for Scientific and Engineering Computing, Italy

23. \ {\bf S.R.S. Varadhan}, Courant Institute, USA

24. \ {\bf X.-G. Wu}, University of Toronto, Canada

\newpage
\centerline{\bf POSTERS }

\vspace{1.5true cm}

 1. \ {\bf E. Aharonov and D. Rothman}, Massachusetts Institute of Technology,
USA

 2. \ {\bf M. G. Ancona}, Los Alamos National Laboratory, USA

 3. \ {\bf  J. Argyris and G. P\"{a}tzold}, University of Stuttgart, Germany

 4. \ {\bf A. Z. Aroguz and A. Taymaz}, Istanbul University, Turkey

 5. \ {\bf J. Bonzani and M. A. Cimaschi}, Politecnico di Torino, Italy

 6. \ {\bf A. Deutsch}, University of Bonn, Germany

 7. \ {\bf U. D'Ortona, D. Salin, J. Banavar, M. Cieplak and R. Robka}, U. P.
M. C., France

 8. \ {\bf B. H.  Elton}, Fujitsu America Inc., USA

 9. \ {\bf P. Grosfils}, Ecole Normale Sup\'{e}rieure, France

 10. \ {\bf D. Gruner. R. Kapral and A. Lawniczak}, University of Toronto,
Canada

 11. \ {\bf S. Hou}, Los Alamos National Laboratory, USA

 12. \ {\bf M.  Krafczyk},  Universt\"{a}t Dortmund, Germany

 13. \ {\bf A. Lemarchand, A. Lesne, A. Perera and M. Moreau},
       Universit\'{e} Pierre et Marie Curie, France

 14. \ {\bf L. S. Lou}, Los Alamos National Laboratory, USA

 15. {\bf G. R. McNamara}, LLNL, USA

 16. \ {\bf M. B. Mineev-Weinstein}, Los Alamos
National Laboratory, USA

 17. \ {\bf K. Molvig}, Exa Corporation, USA

 18. \ {\bf M. Moreau, B. Gaveau, M. Frankowicz and A. Perera},
  Universit\'{e} Pierre et Marie Curie, France

 19. \ {\bf T. Naitoh and M. H. Ernst}, Senshu University, Japan

 20. \ {\bf J. F. Olson and D. Rothman}, Massachusetts Institute of Technology,
USA

 21. \ {\bf Y. H. Qian and  S. Orszag}, Princeton University, USA

 22. \ {\bf P. F. Radkowski}, Radkowski Associates, USA

 23. \ {\bf K. Ravishankar}, SUNY, USA

 24. \ {\bf A. Rovinsky and M. Manzinger}, University of Toronto, Canada

 25. \ {\bf  M. Shevalier and I. Hutcheon}, University of Calgary, Canada

 26. \ {\bf D. Shim, T. Spanos, D. McEhlaney and N. Udey}, University of
Alberta, Canada

 27. \ {\bf B. R. Sutherland and A. E. Jacobs}, University of Toronto, Canada

 28. \ {\bf C. Teixeira}, Exa Corporation, USA

 29. \ {\bf A. Taymaz and A. Z. Aroguz}, Istanbul University, Turkey

 30. \ {\bf R. van der  Sman}, Agrotechnological Research Institute,
Netherlands

 31. \ {\bf B. Voorhees}, Athabasca University, Canada

 32. \ {\bf F. Wang and E.G.D. Cohen}, The Rockefeller University, USA

 33. \ {\bf J. R. Weimar}, Universit\'{e} Libre Bruxelles, Belgium

 34. \ {\bf X.-G. Wu}, University of Toronto, Canada

 35. \ {\bf J. Yepez}, Phillips Laboratory, USA

 36. \ {\bf C. Yu}, University of Tokyo, Japan

\newpage

\bigskip\medskip{ \bf	ABSTRACTS  OF INVITED TALKS}
\vspace{2.0true cm}
\begin{center}
{\bf Large Liquid-Gas Models on 2D and 3D Lattices   } \\
{\bf  C\'ecile Appert and St\'ephane Zaleski } \\

 Laboratoire de Mod\'elisation en M\'ecanique, CNRS,\\
Universit\'e Pierre et Marie Curie, Tour 66,\\
4 Place Jussieu, 75252 Paris Cedex 05, France \\\end{center}

Liquid gas models on a lattice are derived from lattice gas cellular
automata by adding interactions at a distance. These interactions
allow to create a separation of phases. An
interface between a liquid phase  and a gas phase is created
spontaneously. The model is found to obey
the Boltzmann molecular chaos assumptions quite well.
The surface tension and the viscosity
may be predicted using this assumption and an expansion in large
interaction distance. The origin of the non-Galilean
factor will also be discussed. Extensions to three dimensions
were performed recently.
Example of applications are the formation of soap froth and
flow in porous media.

\begin{center}
{ \bf Correlations and Renormalization in Lattice Gas Automata} \\
{\bf B. Boghosian} \\

Thinking Machines Corporation \\
245 First Street \\
Cambridge, MA 02142--1214, USA \\
\end{center}

A method is described for calculating corrections to the usual
Chapman-Enskog analysis of lattice gases due to the buildup of
correlations.  For lattice gases satisfying semi-detailed balance, and
analyzed in the diffusion limit ($\Delta t \sim (\Delta x)^2$), it is
shown that exact renormalized transport coefficients can be calculated
perturbatively by summing a diagrammatic series.  Closed-form
expressions are given for the vertices in these diagrams.  It is shown
that subsets of these diagrams can be easily identified that correspond
to the kinetic ring approximation, or to any truncation of the BBGKY
hierarchy.  This method is applied to several example lattice gases, and
results are shown to be in agreement with numerical experiments.

\newpage

\begin{center}
{ \bf Fluctuation Correlations in Lattice-Gas Automata} \\
{\bf Jean Pierre Boon} \\
Physique Non-Lineare et M\'{e}canique Statistique\\
Universit\'{e} Libre Bruxelles\\
1050 Bruxelles, Belgium
\end{center}

Some aspects of the Statistical Mechanics of Lattice-Gas Automata
will be reviewed with emphasis on density fluctuation correlations.
A comparative analysis of theoretical predictions and simulation
results will be presented.
\vfil

\begin{center}
{\bf Lattice Boltzmann Method and its Application in Computational Biology} \\
{\bf Shiyi Chen } \\

Theoretical Division and Center for Nonlinear Studies \\
Los Alamos National Laboratory \\
Los Alamos, NM 87545, USA
\end{center}

In this talk,  we will present our recent results
of the application of lattice
Boltzmann method in biological systems, including simulation of
cytoskeleton and polymeric fluids.

\vfil

\begin{center}
{ \bf Diffusion, Propagation and Pattern Formulation
 in Lorentz Lattice Gases} \\
    {\bf E. G. D. Cohen and F. Wang} \\
The Rockefeller University,  \\
New York, NY 10021, USA \\
\end{center}

	In a Lorentz lattice gas point particles move, without mutual
interactions, on the bonds of a lattice, whose sites are partly or
fully occupied by scatterers.
	For the strictly deterministic scattering rules for the
particles by the scatterers considered here, the behavior of the
particles differs fundamentally from that for probabilistic scattering
rules.  Various types of diffusive behavior, as well as propagation
and pattern formulation have been observed in computer simulations,
depending on the structure of the lattice, the nature of the
scatterers and their distribution over the lattice. Neither probability
theory nor kinetic theory can account for these phenomena, but a
different theoretical approach, recently initiated by Bunimovich and
Troubetskoy has lead to a number of theorems on the behavior of the
particles on the lattice.
\vfil
\newpage
\begin{center}
{ \bf Polymerization through Heterogeneous Catalysis:
a Reactive Lattice-Gas Automoton Approach} \\
{\bf D. Dab} \\
Physique Non-Lineare et M\'{e}canique Statistique\\
Universit\'{e} Libre Bruxelles\\
1050 Bruxelles, Belgium
\end{center}

We justify the need for a simple microscopic approach to polymerization through
heterogeneous catalysis and we construct a reactive lattice-gas automaton model
which is used to discuss the problem of the broad molecular-weight distribution
observed in real experiments.

\begin{center}
{ \bf Glauber Evolution with Kac Potentials} \\
{\bf A. De Masi} \\

 Dipartimento di Matematica \\
Universit\`{a} dell'Aquila,  \\
67100 l'Aquila, Italy \\
\end{center}

In a typical quenching experiment there are
different regimes with their own
space-time scale and  their characteristic
phenomena.  The early
stage is when the phases develop
emerging from the initial
unstable state.  Clusters of the
thermodynamically stable phases appear,
still relatively small, macroscopically, but
large enough microscopically,
to allow for an accurate statistical
description. In the next stage the clusters move
and this is called the interface dynamics regime.

I study these phenomena in the Glauber spin
flip dynamics with $\pm 1$ valued spins
interacting via a Kac potential. The temperature
is fixed below the critical value and the
initial measure is product with zero average
corresponding to a value of the magnetization
which is thermodynamically unstable.
At times which grow logarithmically
in the scaled length of the Kac interaction, the
early stage of the spinodal decomposition
is observed. We characterize the typical spin
configurations both during the separation and at
the time when the clusters of the two phases
appear. This analysis is strictly related to the
study of the solution of the non local evolution
equation describing the macroscopic behavior of
the model.
The successive motion of the interfaces is also
analized showing that the late stage of the
spinodal decomposition in this isotropic system
with non conserved order parameter, is described
by a motion by mean curvature.
\newpage
\begin{center}
{ \bf  Future Computers and Lattice Methods;
 Multiphase Flows Through Porous Media} \\
{\bf G. D. Doolen} \\

Center for Nonlinear Studies and Theoretical Division
Los Alamos National Laboratory
Los Alamos, NM 87545, USA
\end{center}

Recent developments in electronic devices have shown that lattice
gas emulations could possibly be executed on extremely fast components,
with the possibility that orders of magnitude increase in speed might
be possible. A summary of developments to date will be given along with
some description of progress made in self-assembling computers. Also 3D
lattice Boltzmann calculations of high-resolution multiphase flows through
porous media (10 micron tomographic pore scale data provided by Mobil Oil,
Inc.)
 will be described and videos shown.

\begin{center}
{ \bf Metastability and Pattern Formation
       in Biased Lattice Gases} \\
{\bf  Matthieu  H. Ernst} \\

Institute for Theoretical Physics \\
University of Utrecht \\
 3508 TA Utrecht, The Netherlands \\
\end{center}

Self organization or dynamic phase transitions occur in computer
simulations of a lattice gas with strictly local, but asymmetric
collision dynamics, conserving mass, momentum and lattice symmetries.
A spatially uniform initial state is unstable. At the onset of
instability long wavelength modes drive the system into a
state with long range order. Different domains are not only
characterized by a scalar order parameter (mass density), but also by
one with vector character (momentum density). The structure
factors and spatial correlation functions for the different order
parameters are measured and analyzed in terms of scaling laws.
The state with long range order is highly organized
into moving stable spatial patterns of triangles and parallel
strips of macroscopic size. The onset of instability as well as
the structure of domains and interfaces is well described by mean
field theory.\\
The emphasis at this stage concern the existence and possible
structure and spatial correlations of stationary states or limit
cycles in microscopic models that violate the detailed balance conditions.

\newpage
\begin{center}
{ \bf Lattice BGK Models for Miscible Fluid Flow:
Experiment and Simulations} \\
{\bf E. G. Flekkoy$^{*}$,\vspace{.3ex}
 U. Oxaal$^{**}$, J. Feder$^{**}$,  T. Jossang$^{**}$ } \\

* Center for Advanced Study\\ at The Norwegian
Academy of Science and Letters\\
P.O.Box. 7606 Skillebekk\\
0205 Oslo, Norway\vspace{.6ex} \\
**Departament of Physica \\University of Oslo\\
Box 1048 Blindern\\ 0316 Oslo 3, Norway
\end{center}

We present a comparison between Lattice BGK simulations and a hydrodynamic
dispersion experiment performed in a Hele--Shaw with simple internal
geometry. The aim of this work is to verify the models' to reproduce
quantitatively the experimental results as
well to provide a tool for the extension of these results.
Comparison of the dispersed concentration profiles provides
a test of both the hydrodynamic and the diffusive behavior of the model,
and preliminary results show good agreement between simulations
and experiment.

The two dimensional BGK model simulates, in a simplified way,
the full three dimensional features of the experiment flow.
Although the experiment is performed at a very low Reynolds number,
small effect of nonlinearity is observed.
For steady state flow the model contains a parameter that allows
tuning of the Reynolds  number independently of the flow
velocity, viscosity and system size. This is used to study
 this effect over a wider range of parameters
than what is expermentally accessible.

\begin{center}
{ \bf Lattice Gases and New Emergent Complexity
in Biochemical Systems} \\
{\bf Brosl Hasslacher} \\

 Complex Systems Group Theoretical Division and	\\
Center for Nonlinear Studies \\
Los Alamos National Laboratory \\
Los Alamos, New Mexico 87545, USA \\
\end{center}

Recently, there have been several remarkable developments in new
origins of complexity at the cellular and sub-cellular levels,
which are only accessible at present using large parallel machines
of the CM class. The main biochemical pathway under study is the
glycolytic pathway in the Sel'kov approximation, which is known to
be reliable in laboratory scale glycolytic reactions. We are using
reactive lattice gases, which capture noise effects correctly, to
study diffusion driven instabilities in these reaction-diffusion
systems in several regimes.\vspace{.6ex}

 The first is the pure Turing region
where we observe classic static Turing global symmetry breaking at
the 30 - 500 nanometer scale in ATP concentration, contrary to
previous rough estimates.\vspace{.6ex}

 The second is the Turing-Hopf region
where we see a remarkable range of dynamic ATP concentration
behavior, which so far is analytically unexplored. \vspace{.6ex}

The third is in
the same regimes, but now using strongly perturbed initial conditions
which cause the system to ring in a nonlinear way. \vspace{.6ex}

The latter
excites a remarkable range of complex spatial dynamics that resembles
the actual structures seen in biology, including cell division and
growth of cell walls. This appears to be a new route to chaos. We
feel this is directly relevant at least to the organization and
growth of a large variety of ATP engines, including tubulin and
microtubule dynamics and protein dynamics in general.\vspace{.6ex}

It may also
shed some light on their self-assembly. Similar structures have
recently been seen experimentally, in more limited regimes in other
systems, by both the Texas group under Swinney and the Bordeaux
group under DeKepper. It seems that there are no computational
tools of similar scope available to study these effects. Their
discovery throws into question nearly half century's assumptions
about biochemical dynamics at the sub-micron scale.

\begin{center}
{ \bf Vortex Street and L\'{e}vy Walk} \\
{\bf F. Hayot} \\

Department of Physics \\ Ohio State University\\
Columbus Ohio 43210, USA \\
\end{center}

A vortex street-the so-called von Karman street-is generated
in flow around an infinite cylinder at sufficiently high Reynolds
number. The vortex shedding by the cylinder occurs at a
well defined frequency, and the vortices themselves have a spatial
extent characterized by cylinder size. L\'evy walks correspond to
momentum exchanges over many scales in the flow enveloping the
cylinder. Their intensity and their maximum size model the
presence of some characteristic turbulence in the incoming
flow. One is interested in how this turbulence affects the
vortex street. The L\'evy walk algorithm of lattice gas hydrodynamics
provides a model where the issues of how the scales of
momentum exchanges interfere with the coherent structure
of the vortex street can be investigated.
I will present a number of results and compare them with experiment.

\begin{center}
{ \bf Lattice Gases without Semi-Detailed Balance} \\
{\bf                            M. Henon} \\

                  CNRS, Observatoire de Nice,\\
              BP 229, 06304 Nice Cedex 4, France \\
\end{center}

Is it all right to use lattice gas models without semi-detailed
balance in numerical simulations of fluids ? What is going on when
we do that, and what are the consequences ? In particular, why is
the observed viscosity systematically larger than the theoretical
value computed under the Boltzmann approximation ? In this talk I
will describe a few experiments provoked by these questions.

1. Microscopic correlations have been measured for several
variants of the FCHC 24-velocity lattice gas. They include
correlations between input velocities; between output velocities;
between input and output velocities; between nodes; and between
time steps. Some observed features, but not all, can be explained
as consequences of the criteria used to build the collision
tables.

2. Measurements were also made of the components of the
second-order momentum, which plays a crucial role in connection
with the viscosity.

3. These measurements, taken together, lead to an explanation for
the fact that the observed viscosity is larger than the
theoretical Boltzmann value. An attempt was made to use this
explanation to build a better FCHC collision table;
unfortunately, only a marginal improvement is obtained.

4. When semi-detailed balance is violated, the phase space
accessible to the system as a whole contracts with time, until
eventually a stable subset is reached. This might be relevant to
the observed behavior of the models. In particular it would be of
interest to estimate the time scale of the contraction, and the
size of the final subset. A direct numerical attack seems
impossible in the FCHC case because the phase space is too huge.
Therefore experiments were made with a much simpler lattice, with
one dimension and two or three velocities (in progress at the time
of this writing).

\begin{center}
{ \bf Exact Solutions for the Lattice Boltzmann Equation:
Boundaries and Interfaces} \\
{\bf D. d'Humi\`{e}res} \\

Laboratoire de Physique Statistique \\
24 Rue Lhomond \\
75231 Paris, Cedex 05, France \\
\end{center}

Relaxation Lattice Boltzmann Equation was proposed by \\Higuera [1]
as an improvement of Lattice Gas Models with respect to noise
and ease of implementation for the 3-D models.
It turns out that this scheme has exact solutions for simple flows,
like Couette or Poiseuille ones, both for simple fluid model
and for immiscible ones [2].\\
These exact solutions will be presented along with
their use for a precise characterization of boundary conditions: walls, in
and out flow conditions and interfaces between two fluids.\\

\hangindent=1cm\hangafter=1
[1] F. Higuera and J. Jimenez, ``Boltzmann Approach to Lattice Gas
    Simulations'', {\em Europhys. Lett.} 9, 663 (1989).

\hangindent=1cm\hangafter=1
[2] A.K. Gustensen, D.H. Rothman, S. Zaleski, and G. Zanetti,
    ``Lattice Boltzmann model of immiscible fluids'',
    {\em Phys. Rev.} A43, 107 (1991).
    A.K. Gustensen and D.H. Rothman, ``Microscopic Modeling of Immiscible
    Fluids in Three Dimensions by the Lattice Boltzmann Method'',
    {\em Europhys. Lett.} 18, 157 (1992).
\newpage
\begin{center}
{ \bf CAM-8: A Computer Architecture Based on Cellular Automata} \\
{\bf N. Margolus} \\

Laboratory for Computer	Science \\
Massachusetts Institute of Technology \\
Cambridge, MA 02139, USA \\
\end{center}

CAM-8 is an indefinitely scalable multiprocessor optimized for
spatially fine-grained, discrete modeling of physical systems---such
as lattice-gas simulations of fluid flows.  With an amount and kind of
hardware comparable to that in an inexpensive workstation (64
Megabytes of conventional DRAM and 2 Megabytes of cache-grade SRAM,
all running with a 25 MHz clock) our small-scale prototype already
performs a wide range of such simulations at speeds comparable to the
best numbers published for any commercial machine.  Machines orders of
magnitude bigger and proportionately faster can be built immediately,
using the existing (working) chips.  This kind of computing power has
never before been available for spatially fine-grained modeling: CAM-8
makes a new band of the computational spectrum effectively accessible.

\begin{center}
{ \bf The Discrete Boltzmann Equation for Gases
with Bi-Molecular or Dissociation-Recombination Reactions} \\
{\bf Roberto Monaco} \\

Department of Mathematics \\
University of Genoa \\
Via L. B. Alberti 4 - 16132 Genova-Italy \\\end{center}

In recent years, new interest has been devoted to real gas effects due
chemical reactions.  In fact, as well documented by the Proceedings book of an
IUTAM Symposium [1], in the near future, several space missions are foreseen in
the
atmosphere of celestial bodies of our planetary system where chemical reactions
play a major role.  In this physical situation, problems including combustion
[2],
catalytic gas-surface interactions [3], and onset of shock-waves and of
detonation
waves [4] must be investigated both at the macroscopic and the microscopic
level.

A physical mathematical tool which seems to be promising, for the
representation
at a microscopic level of a chemically reacting gas mixture, is the Discrete
Boltzmann Equation (DBE), see Chap. 6 of book [5].

This paper has the aim of proposing two kinetic models of the DBE, the first
representing gases undergoing bi-molecular reactions, the second representing
gases with
dissociation-recombination reactions.

The first model considers a mixture of four gases undergoing
elastic collisions and reactions of the type

$A  + B  \rightleftharpoons C + D$ .

The second is related to reactions for a diatomic gas $A_2$ , i.e.

$A_2 + M \rightleftharpoons A + A^* + M$

and considers binary and triple elastic collisions as well.  In the second
equation, $A$ are
atoms and $A^*$ atoms at a higher energy level.  M is a catalyzer which can be
either a molecule $A_2$, either an atom $A$ or
either a high energetic atom $A^*$.

For the first model we choose a Broadwell-type discretization of velocity, i.e.

${\bf v}_i^M  = \mu_M c {\bf e}_i , ~ ~     i = 1,..,6,  ~~   M = A,B,C,D, $

$\mu_M$ being the mass ratios $m_M/m_A$; $c$ is a reference speed and ${\bf
e}_i$ are
unit vectors oriented towards the positive and negative directions of the axes
of
a spatial orthogonal frame.

The second model is derived in the plane.  In detail, we consider a square
centered in the origin of an orthogonal frame.  Then the velocities of atoms
$A$
are directed in the positive and negative directions of the x- and y-axes.
Molecules have the same velocities as atoms but with a smaller speed.  Atoms
$A*$
finally have velocities in the direction of the four vertices of the square.
In
symbols

$A :	{\bf v}_i = c{\bf e}_i, ~~    i = 1,...,4 , ~~    {\bf e}_i =
\{i,j,-i,-j\}$

$A_2 :	{\bf v}_i  = \mu {\bf e}_ic,  ~~    \mu = 1/2m_A$

$A* :	{\bf v^*}_i = c{\bf i}_i, ~~   {\bf i}_i  =
\{{\sqrt 2}({\bf i}+{\bf j}),
{\sqrt 2}(-{\bf i}+{\bf j}),
{\sqrt 2}(-{\bf i}-{\bf j}),
{\sqrt 2}( {\bf i}-{\bf j})\},$

where $m_A$  is the mass of the atom and $c$ again a reference speed.
	For both models, we write the kinetic equations, following the hypothesis:\\
1)	Elastic and chemical interactions are characterized by different collision
frequencies.\\
2)	Each collision  preserves mass, momentum and energy (including chemical link
energy).\\
	In the present paper, starting from this point, we derive for both models

	$\bullet$ the space of collisional invariants and the conservation equations

	$\bullet$ the H-theorems, the thermodynamical equilibrium conditions and the
mass-action laws

	$\bullet$ the Euler equations.

	Moreover for the first model we obtain

	$\bullet$ the Navier-Stokes equations and the transport coefficients.

	Finally for the second model we study by numerical simulations

	$\bullet$ the effects of the dissociation rate on the Riemann problem for the
shock-waves onset.

References\\
\noindent
1. {\em	Aerothermochemistry of Spacecraft and Associated Hypersonic Flows}, Ed.
R.
Brun, IUTAM-Marseille, 1992.\\
2. {\em	Recent Advances in Combustion Modeling}, Ed. B. Larrouturou, World Sci.
Pub.,
Singapore, London 1991.\\
3. C. Bruno, in {\em Fluid Dynamical Aspects of Combustion Theory}, Eds. M.
Onofri and
A. Tesei, {\em Pitman Series in Math.}, Longman Sci. and Tech., J. Wiley, New
York,
1992, p. 329.\\
4. A. K. Kapila, ibidem, p. 143\\
5. R. Monaco and L. Preziosi, {\em Fluid Dynamic Applications of the Discrete
Boltzmann Equation}, World Sci. Pub., Singapore, London, 1991.

\begin{center}
{ \bf Collision Rules for Two-Dimensional Hydrodynamics} \\
{\bf Alain Noullez} \\

  Mech. \& Aerospace Eng. Dept.\\
Room D-414 E-Quad, Olden Street,\\  Princeton University\\
  Princeton, NJ-08544, U.S.A.
 \end{center}

I investigate the design of optimized collision rules for the 2-D FHP
lattice gas model for minimal viscosity.  It is shown that for these models,
the optimization reduces to a geometrical matching problem whose solution
can be obtained explicitly for any number of static particles.  This
construction is used to obtain models which are more efficient than the FCHC
model for two-dimensional hydrodynamics.  The effect of relaxing the constraint
of semi-detailed balance and its effect on the equilibrium properties of
these models is also investigated.

\begin{center}
{ \bf Lattice Boltzmann Simulations of Pattern Formation
Reaction-Diffusion Systems}  \\
{\bf S. Ponce-Dawson} \\

Center for Nonlinear Studies and Theoretical Division \\
Los Alamos National Laboratory \\
Los Alamos, NM 87545, USA \\
\end{center}

I will describe a lattice Boltzmann model for
reaction-diffusion equations that has been developed recently [1].
In particular, I will  discuss the ability of the scheme
to simulate the formation of patterns in these systems.
I will analyze the appearance of these patterns due to the
Turing instability in cases with and without
convection of the reactants, the differential flow induced instability
and the interaction among them.\\
 1. S. Ponce-Dawson, S. Chen, and G. Doolen, {\bf J. Chem. Phys.}
{\bf 98}, 1514 (1993)

\begin{center}
{\bf Critical Fluctuations in Spin Systems} \\
{\bf Errico Presutti} \\

Dipartimento di Matematica \\
II Universit\`{a} degli Studi di Roma \\
Via Fontanile di Carcaricola \\
00133 Roma, Italy \\
\end{center}

I consider an Ising spin system on the lattice \
${\bf Z}^{d}$ with ferromagnetic interactions given
by a Kac potential.  The problem that I want to
discuss concerns the structure of the
macroscopic fluctuations of the magnetization at
the critical temperature, in the scaling limit
when the range of the Kac potential becomes
infinite.  In particular the question is when the
limiting process is non Gaussian  and, in that
case, a control of the ultra-violet
divergencies in the discrete approximation.

Gibbsian equilibrium and non equilibrium Glauber
spin flip dynamics are considered.  In d=1, it is
proven that, on a given space time scaling, the
fluctuations process, suitably renormalized,
converges to a non linear Ginzburg Landau
equation with noise.  In particular therefore we
have that a discrete model for the
stochastic quantization of the anharmonic
oscillator is the Ising Glauber
dynamics with Kac potentials at the critical
temperature. The convergence
result is derived by comparison of the true
process with another spin flip evolution, known
as the voter model.  It is shown that while the
voter model converges to the Gaussian process
given by the linear Ginzburg Landau equation with
noise, the density (Radon Nykodim derivative)
with respect to the voter model converges to the
density of the non linear Ginzburg Landau
equation with respect to the linear one.

In d=2, I consider the equilibrium fluctuations
and prove the following.  If the inverse
critical temperature, $\beta = 1$, is
approached from below (with suitable speed)
then the limiting process is Gaussian with an
extra mass related to the Wick regularization
term.  This leads to the conjecture that
the non linear evolution can
be actually derived by approaching the
critical temperature with some
different speeds.  The dynamical aspects of
this procedure are also discussed.

\begin{center}
{ \bf Lattice Gas Transport in Porous Media} \\
{\bf R. Rechtman$^{*}$ A. Salcido$^{**}$} \\

$*$ Depto. de F'isica, Facultad de Ciencias, \\
UNAM, Apdo. Postal 70-542		       \\
 04510, M'exico D.F., M'exico			 \\
$**$ Instituto de Investigaciones El'ectricas,	   \\
 Apdo. Postal. 475\\
   6200, Cuernavaca Mor., Mexico \\
\end{center}

 A nine velocities lattice gas is used for computer simulations
of 2-d flows through random porous media. We study self diffusion,
the dependence of the diffusion coefficient on temperature and the rate
flow.

\begin{center}
{ \bf Surface Tension and Immiscible Lattice Gases} \\
{\bf Daniel H. Rothman} \\

Laboratoire de Physique Statistique\\
Ecole Normale Sup\'erieure\\
24 rue Lhomond\\
75005 Paris, France \\
\end{center}

The talk begins with a review of immiscible lattice-gas models and related
lattice-Boltzmann models of immiscible fluids.
Recent applications are briefly reviewed, with some emphasis on
the problem of multiphase flow through porous media.
I then describe a theoretical calculation, based on
a Boltzmann approximation, of the surface tension in immiscible
lattice-gas models.
Among other results, the calculation shows a phase transition
from a mixed state to a phase-separated state via the existence
of a non-zero surface tension above a critical particle
density $d_c \approx 0.2$.
The accord between the theoretical predictions and empirical
measurements of the surface tension are qualitatively, but not
quantitatively, good.  Errors are due to the neglect of correlations,
which appear to strongly influence the magnitude of the surface
tension at high particle densities.
\newpage
\begin{center}
{ \bf Two and Three-Dimensional Simulations of Rayleigh-B\'{e}nard
Turbulence with Lattice Boltzmann Method} \\
{\bf S. Succi$^{*}$, F. Massaioli$^{**}$,
 R. Benzi$^{**}$, R. Tripiccione$^{***}$} \\

*   IBM European Center for Scientific and Engineering Computing\\
              00141 Roma, Italy			\\

** Physics Department, University of Roma, \\
Via Ricerca Scientifica\\ 00133 Roma, Italy \\

*** INFN Sezione di Pisa,\\ Pisa, Italy

\end{center}

Numerical simulations of Rayleigh-B\'{e}nard Convection in two and three
dimensions using the Lattice Boltzmann method are presented.\\

In particular, a number of new highlights related to the onset
of soft-to-hard turbulence transition, probability distribution
functions and scaling laws in thermal turbulence will be discussed.

\begin{center}
{ \bf  Asymmetric Mean Zero Random Walk with
       Exclusion: Self Diffusion} \\
{\bf  S.R.S.Varadhan} \\

Courant Institute, New York University \\
251, Mercer Street, \\
New York, NY 10012, USA \\
\end{center}

We study the motion of a tagged particle in equilibrium in the case of
an asymmetric mean zero random walk with exclusion. \\We establish
convergence to Brownian motion under the usual diffusive scaling of
space and time. The non-reversibility of the model causes problems
and the usual methods have to be modified in order to establish
the result.

\begin{center}
{ \bf Reactive Lattice-Gas Automata and Chemical Chaos} \\
{\bf Xiao-Guang Wu} \\

Chemical Physics Theory Group \\
 Department of Chemistry \\
 University of Toronto,\\
Toronto M5S 1A1, Canada \\
\end{center}

An overview of the methods used to construct lattice-gas cellular automata
for multi-component chemically reacting systems will be given. As an example
of the application and utility of this method, a mesoscopic model
of  the  Willamowski-R\"{o}ssler  reaction,  a three-variable system whose
mean-field  rate  law  gives  rise  to  a  strange  attractor, will be
constructed.
The effects of fluctuations on the dynamics in the regime
where ``deterministic" chaos exists will be studied and the validity of
the mean-field description will be examined.
\newpage

\bigskip\medskip{ \bf	ABSTRACTS OF POSTERS}
\vspace{2.0true cm}

\begin{center}
{ \bf A New Numerical Model of Non-Newtonian Fluids} \\
{\bf Einat Aharonov and Daniel H. Rothman} \\

Department of Earth, Atmospheric, and Planetary Sciences\\
Massachusetts Institute of Technology\\
Cambridge, MA \ 02139 USA \\
 \end{center}

We introduce a new numerical model of non-Newtonian fluids based on
an idealized microscopic kinetic theory.
Specifically, we simulate a Boltzmann equation in which the
collision dynamics are dependent on the local instantaneous strain.
We use the new model to study flow through porous
media, a problem that has applications in the flow of molten magma
through the mantle and flow of water and contaminants through soil,
and find that flux is related to force by a simple scaling law.
\vspace
{1.0true cm}

\begin{center}
{ \bf Lattice-Boltzmann Methods for Simulating Semi-Classical
Transport Phenomena in Semiconductors} \\
{\bf M. G. Ancona} \\

Center for Nonlinear Studies, MS-B258 \\
Los Alamos National Laboratory \\
Los Alamos, NM  87545 \\
ancona@goshawk.lanl.gov \\
\end{center}

We discuss two different lattice-Boltzmann schemes for solving macroscopic PDEs
describing semi-classical electron transport phenomena in semiconductor
devices.
The first scheme solves the simple yet widely-used equations of the
diffusion-drift transport description.  In this we include both electron and
hole transport with realistic mobility models for both carrier types.  The
second scheme supplements the first by including effects of electron inertia,
an
inclusion which is important for accurate simulation of high-frequency devices.
For both of these schemes we solve the coupled electrostatics problem using a
lattice-Boltzmann -based relaxation procedure thereby maintaining an overall
consistency of method.  The poster will describe each of these schemes in some
detail, give evaluations of them on numerical grounds and present results
pertaining to silicon and silicon-germanium field-effect transistors obtained
using them on a massively-parallel computer (CM-200/CM-5).

\newpage
\begin{center}
{ \bf Lattice-Boltzmann Simulation of Thermal Turbulence} \\
{\bf John Argyris and Gerald P\"{a}tzold} \\

Institute for Computer Applications \\
University of Stuttgart \\
Pfaffenwaldrin g 27 \\
W-7000 Stuttgart 80, FRG \\
\end{center}

In this contribution, we present our lattice-Boltzmann simulations of
convective flow.  Our method is suited to two- and three-dimensional
calculations.  We use the lattice geometry presented in reference [4].  Our
treatment of convective flow in the Boussinesq approximation relies on the
analogy between mass and heat transport.  We have already used the same
principle in corresponding lattice-gas simulations, cf. references [1] and [2].
The collision term in the lattice-Boltzmann equation is represented by a
two-relaxation-times model, corresponding to the two transport coefficients
in the macroscopic equations.  To incorporate a buoyancy force, a slight
perturbation is added.  We can choose the relaxation times in the domain of the
known stability limits, i.e. the transport coefficients must remain positive.

Right from the beginning, we have implemented our lattice-Boltzmann model on a
parallel computer with SIMD architecture.  The machine, a MasPar Mp 1216,
possesses a data parallel unit consisting of 16384 processor elements (PE),
arranged in an array of size 128 x 128.  Each PE is endowed with 16 kByte local
memory.  The programming has been done using the MasPar implementation of the
FORTRAN 90 language.  For the performance of the code, it is of importance that
the communication between the PEs is done by the fast ``xnet''-mechanism of the
machine, which connects eight nearest neighbors to each processor.

As an application, we want to present our first simulations of thermal
turbulence in Rayleigh-B\'enard convection.  To verify the lattice-Boltzmann
algorithm in this regime, we oriented our work to the simulations presented in
reference [5] (see also [7]).  So the calculations are performed for a
convection cell of unit aspect ratio.  The Prandtl number has a value of 7 and
the Rayleigh number varies between $10^5$ and $10^7$.  Also in accordance with
experimental findings (see references [3] and [8]), we observe waves and plumes
which develop from the thermal boundary layer.  Currently, closer analysis of
the simulation data is in progress.  Here, we will use methods of wavelet
analysis, in the spirit of reference [6].

References \newline
[1]  J. Argyris, G. Faust, and G. P\"{a}tzold.  Cellular Automata and
Convection.
Poster presented at the  {\em Second World Congress on Computational
Mechanics},
August 27-31, 1990, Stuttgart, FRG. \newline
[2]  J. Argyris and G. P\"{a}tzold.  Two-Dimensional Lattice Gases for Regular
Binary Mixtures.  In:  Proceedings of the Fourth European Turbulence
Conference,
special issue of {\em Applied Scientific Research}, edited by F. T. M.
Nieuwstadt (to
be published in spring 1993). \newline
[3]  B. Castaing et al. Scaling of Hard Thermal Turbulence in Rayleigh-B\'enard
Convection.  {\em J. Fluid Mech.} 204, 1-30 (1989). \newline
[4]  S. Chen et al. Lattice Boltzmann Computational Fluid Dynamics in Three
Dimensions.   {\em J. Stat. Phys.} 68 (3/4), 379-400 (1992). \newline
[5]  E. E. De Luca et al. Numerical Simulations of Soft and Hard Turbulence:
Preliminary Results for Two-Dimensional Convection.  {\em Phys. Rev. Lett.} 64
(20),
2370-2373 (1990). \newline
[6]  C. Meneveau.  Analysis of Turbulence in the Orthonormal Wavelet
Representation.  {\em J. Fluid Mech.} 232, 469-520 (1991). \newline
[7]  L. Sirovich et al. Simulations of Turbulent Thermal Convection.  {\em
Phys.
Fluids} A 1 (12), 1911-1914 (1989). \newline
[8]  G. Zocchi et al. Coherent Structures in Turbulent Convection, an
Experimental Study.  {\em Physica} A 168, 387-407 (1990). \newline

\begin{center}
{ \bf Pattern Formation Simulation of Certain Chemical Reactions} \\
{\bf Ayse Zehra Aroguz$^1$ and Adnan Taymaz$^2$} \\

$^1$Chemistry Department, Engineering Faculty, Avcilar Campus \\
$^2$Physics Department, Science Faculty, Vezneciler Campus \\
Istanbul University, 34459 Istanbul, Turkey \\
\end{center}

The computer simulation technique of molecular dynamics is briefly reviewed,
which involves simulation of chemical reaction function of molecules on a
compute from a few hundred small molecules such as $H_2O$ or $N_2$ over a few
picoseconds to larger macro-molecules.  The aim of this presentation is two
fold; first to introduce a mathematical procedure which enhances visual
separation of individual band of a molecule, second to deal with the
investigation of certain molecular patterns of practical interest at atomic
level.  The validity of reaction function simulation has been dealt by
comparison with experimental results.

\begin{center}
{ \bf A Discrete Kinetic Model with Chemical
Reactions of Type $A + A \to B$} \\
{\bf Ida Bonzani$^1$ and M. Antonietta Cimaschi$^2$}  \\

$^1$Department of Mathematics, Politecnico di Torino \\
$^2$Department of Mathematics, University of Genova \\
\end{center}

Discrete kinetic theory for gases undergoing to chemical reactions have been
introduced in [1].  As known gas mixtures with bi-molecular dissociation and
recombination reactions are considered both in aerospatial engineering [2] and
combustion theory [3].  The complete Boltzmann equation extended to chemically
reacting gases leads to a complicated mathematical structure in terms of
integro-differential equations.  Discrete models seems to be more convenient in
view of fluid dynamic applications and particularly in cellular automata
research.

In the present paper, we consider a binary gas mixture, undergoing to the
following chemical reaction: $A + A \to B$

Each gas species may experience binary elastic collisions too, between
particles
of the same species or opposite species; both chemical and elastic interactions
preserve mass and momentum.

The set of admissible velocities for the gas particles is given by the
following
sixteen planar vectors

			${\bf v}_{2k-1}^A = c{\bf e}_k, ~~	       (k = 1,...,4)$

		        ${\bf v}_{2k}^A  = {\bf v}_{2k-1}^A + {\bf v}_{2k+1}^A $

			${\bf v}_i^B = 2{\bf v}_i^A, ~~	(i = 1,...,8)$

where ${\bf e}_k = \{{\bf i}, {\bf j}, {\bf -i}, {\bf -j}\}$ are vectors in the
fixed frame
$(O, {\bf i}, {\bf j})$ and $c$ is a reference speed.
If the number densities related to each velocity are

			$N_i^A = N_i^A (t, x),$

			$N_i^B = N_i^B (t, x),~~  x \in R^2$,

the resulting kinetic equations are expressed in the form

	$ {\partial N_i^A \over \partial t} + {\bf v}_i^A \cdot \nabla_x N_i^A =
 J_i^A (N^A, N^B)  - R_i^A(N^A)$

	$ {\partial N_i^B \over \partial t }+ {\bf v}_i^B \cdot \nabla_x N_i^B =
 J_i^B (N^A, N^B)  + R_i^B(N^A)$

where the terms $J_i$  are due to elastic collisions and $R_i$  to chemical
reactions.

Mathematical properties of the model and physical implications are then
considered in the paper, together with some numerical simulations where classic
flows in  bounded and unbounded domains are proposed and visualized.

References\\
\noindent
1. R. Monaco and L. Preziosi, {\em Fluid Dynamic Applications of the Discrete
Boltzmann Equation}, World Sci. Pub. (1991). \\
2. IUTAM Symp. {\em Aerothermochemistry of Space Crafts}, Ed. R. Brun,
Marseille
(1992). \\
3. {\em Fluid Dynamical Aspects of Combustion Theory}, Eds. Onofri M. and Tesei
A.,
Pitman Series of Math., J. Wiley, New York (1992). \\

\begin{center}
{ \bf Towards Analyzing Complex Swarming Patterns in
Biological Systems with the Help of Lattice-Gas Cellular Automata} \\
{\bf Andreas Deutsch} \\

Biologie \\
University of Bonn \\
\end{center}

Cellular automaton models have been successfully applied to biological
phenomena, in particular to the formation of morphogenetic patterns [2,3,7].
Nevertheless, an important problem of cellular automata lies in their limiting
behavior, i.e. the consistency of the discrete automaton formulation with
continuous models written in the language of partial differential equations.
 From lattice-gas theory many examples are known in which this problem has been
solved [4].  While the core of a (physical) lattice gas consists of rotation,
propagation and collision operators mimicking the dynamics of `elementary
particles', chemical dynamics (e.g. of reaction-diffusion type) can be modeled
by introducing additional `reactive' operators [6].

Here, we address the problem of swarming behavior in biological systems.
Swarming phenomena are not limited to organisms but may also occur on a
cellular
level.  In any case, certain interactions between cells or organisms are
responsible for the origin of typical swarming patterns.  For example, certain
salamander larvae exhibit horizontal and vertical stripe patterns of pigment
cells which arise from migration, i.e. swarming of embryonic cells within the
fibrous network given by the extracellular matrix [5].

A biological `lattice-gas' cellular automaton is developed that is based on a
piecewise straight random walk with constant speed of organisms (or cells).
The
temporal dynamics of the automaton is defined by rotation and propagation
operators together with a `biological' operator which describes local
interactions of organisms (or cells).  The microdynamic equations are given and
from these a continuous transport equation of Boltzmann type (see [2]) may be
deduced.  Possible applications are the swarming of the myxobacteria and ants,
the contractile motion of actin-myosin fibrils, as well as the formation of
aggregation patterns of Dictyostelium and the mentioned pigment cell patterns
in
salamander larvae.  Furthermore, examples of simulations are shown.

References

\noindent
1.	Alt, W., and Pfistner, B. (1990) in : Alt, W.
 and Hoffmann, G. (eds.) {\em Biological Motion},  Springer, Berlin,
Heidelherg. \\
2.	Deutsch, A. (1992) In:  Rensing, L. (ed.) {\em Oscillations and
Morphogensis},
Marcel Dekker, New York. \\
3.	Deutsch, A., Dress, A., and Rensing, L., (1993) To appear in: {\em
Mechanisms of
Dev.} \\
4.	Doolen, G., Frisch. U., Hasslacher, B., Orszag, S., and Wolfram, S., (1990)
{\em Lattice Gas Methods for Partial Differential Equations},  Addison-Wesley,
Redwood
City, New York. \\
5.	Epperlein, H.-H., and Lofberg, J. (1990) The Development of the Larval
Pigment
Patterns in Triturus Alpestris and Ambystoma Mexicanum,  Springer, Berlin. \\
6.	Kapral, R., Lawniczak, A., and Masiar, P. (1992) {\em J. Chem. Phys.} 96:
2762-2776. \\
7.	Young, D. A. (1984) A Local Activator-Inhibitor Model of Vertebrate Skin
Patterns. {\em  Math. Biosciences} 72: 51-58. \\

\begin{center}
{\bf  Study of Rayleigh-B\'{e}nard Cells with a Cellular Automaton} \\
{\bf U. D'Ortona, D. Salin, J. Banavar$^{(1)}$, M. Cieplak$^{(2)}$, R.
Rybka$^{(3)}$} \\

               A.O.M.C.- U.P.M.C. case 78, Tour 13,\\
           4, pl Jussieu, 75252 Paris Cedex 05, France
\end{center}

We modify a 2-dimensional Boltzmann Cellular Automaton to
mimic the advection of a passive or an active contaminant.
If the contaminant is temperature, the gravity effect is
achieved by slightly changing the flow depending on the
temperature, leading to the formation of Rayleigh-B\'{e}nard
convection cells. We show that the implementation of conduction
is necessary to obtain a regular pattern of cells. The same
study is realized in a porous medium. The porous medium is obtain
by randomly putting unaccessible sites in the lattice. In this
case the instability is the Rayleigh-Darcy's one.
\noindent
(1) Penn-State Univ. 104 Davey Laboratory, University Park,

\hspace{.6true cm} PA16802, USA \\
(2) Institute of Physics, Polish Academy of Sciences,

\hspace{.6true cm} 02-668 Warsaw, Poland	 \\
(3) Institute
of Geophysics, Polish Academy of Sciences,  01-452 Warsaw, Poland \\
\vspace{1.0true cm}

\newpage
\begin{center}
{\bf Comparisons of Lattice Boltzmann Methods with a Finite Difference Method
     for a 2-D Burgers Equation} \\
{\bf Bracy H. Elton}\\

Computational Research Division \\
Fujitsu America, Inc.         \\  
3055 Orchard Drive          \\    
San Jose, CA  95134-2022,  USA
\end{center}

We look at three lattice Boltzmann methods and an analogous second-order
finite difference method for solving a two-dimensional scalar, viscous
Burgers equation with periodic boundary conditions.  Specifically, we
compare the four methods in terms of convergence attributes, including
domain of monotonicity, order of convergence, and absolute errors, and
in terms of their performance characteristics, including performance and
timing measurements and memory requirements, on Fujitsu supercomputers.
\vspace{1.0true cm}

\begin{center}
{ \bf Latice Gas Automaton Model for the Coupling between Internal and
Translational Modes} \\
{\bf Patrick Grosfils} \\

Laboratoire de Physique Statistique \\
Ecole Normale Sup\'{e}rieure \\
24 Rue Lhomond \\
75231 Paris, Cedex 05, France \\
\end{center}

We consider a 2-D lattice gas model is order to introduce a relaxation time
for the rest particles. A computation of the spectrum of density fluctuations
shows dispersion for acoustic waves together with a ``Mountain" mode.
\vspace{1.0true cm}

\begin{center}
{ \bf Nucleation, Domain Growth and Fluctuations in a Bistable Chemical System
   } \\
{\bf Daniel Gruner$^{*}$, Raymond Kapral$^{*}$ and
Anna Lawniczak$^{**}$} \\
* Department of Chemistry\\ University of Toronto\\ Toronto, Ontario,
	   Canada M5S 1A1    \vspace{1.5ex}
** Department of Mathematics and Statistics\\
University of Guelph\\
Guelph, Ontario, Canada N1G 2W1
\end{center}

Phase separation and nucleation processes are investigated for a
bistable chemical system.  The study utilizes a reactive lattice-gas
cellular automaton model to provide a mesoscopic description of the
dynamics.  Simulations of steady-state structure, wave propagation, and
critical nucleus size using this model are compared with results based
on the deterministic equations of motion.  The dynamic structure factor
is computed for evolution from the unstable state and the effects of
correlations are examined for early and late times.  The study provides
insight into these processes in a fluctuating, extended medium and also
provides a test of the ability of the reactive lattice-gas method to
describe the fluctuations in the system.

\begin{center}
{ \bf Comparisons between the Lattice Boltzmann Method and Traditional CFD
Methods
for Two-Dimensional Cavity Flow} \\
{\bf Shuling Hou} \\
Center For Nonlinear Studies, Los Alamos National Laboratory
\end{center}

Despite some applications of the Lattice Boltzmann methods in hydrodynamics
and other fields, quantitative studies of the method have been limited.
In this work, Lattice Boltzmann BGK model (LBBGK) has been applied to
the 2D driven cavity flows for Reynolds numbers up to 10,000. Detailed
comparisons
between the LBBGK method and traditional methods are performed
and show excellent agreement. Also, the compressibility error of
LBBGK methods and their convergence rates are discussed.
\begin{center}

{\bf Coupling of Lattice-Gas and Finite Element-Methods} \\
{\bf Manfred Krafczyk} \\

Lehretuhl NMI, \ GB II\\	 Universt\"{a}t Dortmund \\
August Schmidt Str. 8b \\4600 Dortmund 50, Germany \\
\end{center}

In the last decade there has been a very successful development of Lattice
Gas (LG)
algorithms for simulation of flow-problems and related topics parallel to
the refinement of ``classical" methods like Finite Differences,
Finite Elements (FE) and spectral methods.
Due to their inherent structural differences LG- and FE-algorithms
show specific advantages and disadvantages when imposing them on
specific parts of e.g. multiphase-flow-problems governed by the incompressible
Navier-Stokes-Equations. The main difference can be recognized in the fact
that LG-methods are strictly local algorithms while FE-methods proceed
(typically) in a non-local way.
While analyzing problems where both local and non-local interactions
are equally important, it is evidently desirable to couple both algorithms
in order to gain the advantages of both formalisms so that the efficiency of
simulations is increased.
In order to demonstrate the improvement when using a ``mixed" algorithm we
implemented the so-called Immiscible Lattice Gas (ILG, Gunstensen \&
Rothman, '91)
in its Galilean invariant form and coupled it with a FE-program for field
computations. We show convergence-acceleration which steams
directly from the physically motivated coupling of FE + ILG - Algorithms.

\begin{center}

{ \bf Fractal Character of a Chemical Wave Front} \\
{\bf  A. Lemarchand, A. Lesne, A. Perera, and M. Moreau}

Laboratoire de Physique Th\'eorique des Liquides \\
Universit\'e Pierre et Marie Curie \\
4, Place Jussieu, \\
75252 PARIS Cedex 05, FRANCE

{ \bf M. Mareschal} \\
D\'epartement de Chimie-Physique \\
Universit\'e Libre de Bruxelles \\
Campus Plaine, Bvd du Triomphe \\
1050 BRUXELLES, BELGIUM \\
\end{center}

A reactive lattice gas cellular automaton model is used to simulate a chemical
wave front propagating in  a two-dimensional (2-D) medium.  The corresponding
macroscopic description is given by a reaction-diffusion equation first studied
by Fisher and Kolmogorov, Petrovsky, Piskunov in the 1-D case.  The computed
value of the front propagation velocity agrees with the 1D macroscopic value.

On the contrary, the front width is half the predicted 1-D value.  This result
is explained by the fractal character of the interface.  The fractal structure,
described through several fractal dimensions, is shown to be independent of the
reaction and diffusion parameter values.

\begin{center}
{\bf Building Correct and Stable Modeles for Lattice Boltzmann Hydrodynamics}
\\
 {\bf Guy R. McNamara} \\

LLNL, L-540\\
P.O.Box 808\\
Livermore, CA 94550, USA \\
\end{center}

The Lattice Boltzmann (LB) method of
computational fluid dynamics has shown
particular promise for modeling systems
involving complex boundaries or
multiphase fluids, and has proven suitable for
modeling high Reynolds
number flow.  The original LB models suffered from defects
inherited from
their lattice gas ancestors, but these may be overcome by
modifying the
model's equilibrium mass distribution.  We describe a method of
constructing LB models which makes explicit the lattice
symmetries required
for correct hydrodynamics.  This methodology may be employed to
quickly and
mechanically generate collision operators, with or without energy
conservation, for a variety of lattices.  The LB models so
constructed may
not exhibit numerical stability in the limit of small transport
coefficients, but in some cases this difficulty may be resolved by
introducing additional lattice velocities and appealing to the
 principle of
entropy maximization to extend the equilibrium distribution to the
augmented velocity set.

\begin{center}
{ \bf Multidimensional Pattern Formation Has an Infinite Number of Constants in
Motion } \\
{\bf Mark B. Mineev-Weinstein} \\

Center for Nonlinear Studies \\
Los Alamos National Laboratory \\
Los Alamos, NM  87545 \\
\end{center}

Extending our previous work on 2D growth for the Laplace equation we study here
multidimensional  growth for arbitrary elliptic equations, describing
inhomogeneous and anisotropic pattern formations processes.  We find that these
nonlinear processes are governed by an infinite number of conservation laws.
Moreover, in many cases all dynamics of the interface can be reduced to the
linear time-dependence of only one ``moment'' $M_0$ which corresponds to the
changing
volume while all higher moments, M1, are constant in time.  These moments have
a purely geometrical nature, and thus carry information about the moving shape.
These conserved quantities are interpreted as coefficients of the multipole
expansion of the Newtonian potential created by the mass uniformly occupying
the
domain enclosing the moving interface.  Thus the question of how to recover the
moving shape using these conserved quantities is reduced to the classical
inverse potential problem of reconstructing the shape of a body from its
exterior gravitational potential.  Our results also suggest the possibility of
controlling a moving interface by appropriate varying the location and strength
of sources and sinks.

\begin{center}
{ \bf A New Class of Nonsingular Exact Solutions for Laplacian Pattern
Formation } \\
{\bf Mark B. Mineev-Weinstein} \\

Center for Nonlinear Studies \\
Los Alamos National Laboratory \\
Los Alamos, NM  87545 \\
\end{center}

We present a new class of "N-finger-like" exact solutions for the so-called
Laplacian Growth Equation  describing the zero-surface tension limit of a
variety of 2D pattern formation problems.  We prove that, contrary to the
typical situation in the zero-surface tension limit, these solutions are free
of
finite-time singularities (i.e. they do not develop cusps in a finite time). In
the long-term asymptotics the moving interface consists of N separated fingers.
This evolution from a quite arbitrary initial interface resembles the N-soliton
solution of classical integrable PDE's such as KdV, NLS, etc.

\begin{center}
{ \bf  A New Technology for Fluid Simulation} \\
{\bf Kim Molvig} \\

Exa Corporation\\
125 Cambridge Park Drive\\
Cambridge, MA. 02140, USA
\end{center}

Huge increases in computing power are known to be needed for fluid flow
simulation. The complexity of flow around an automobile or other object moving
at a realistic speeds involves so many degrees of freedom that present methods
and technologies cannot begin to provide enough computational power for
accurate simulation.  Exa Corporation has developed an algorithm and
architectural support that together comprise a new, inherently scalable
technology for fluid simulation which promises dramatic improvements in
computational power - from a workstation-sized engine delivering 50x the fluids
simulation performance of a Cray supercomputer initially to a
PetaFlops-equivalent server ultimately.  Such power is required  to accurately
model all the scales of motion for realistic flow speeds; five of the ten Grand
Challenge problems can be directly addressed with this technology.  This method
represents a major advance in Lattice Gas theory - all discreteness artifacts
have been removed.  The system behaves as though the underlying lattice were
actually erased from the dynamics.  The algorithm is fundamentally more
accurate than discretized approximations to the Navier-Stokes equations, it is
easier to apply as it uses a simple rectilinear grid, it is computationally
much more efficient than methods based on floating point arithmetic, and it is
inherently parallel.

The technology is based on a direct representation of physical fluids as a
three-dimensional ``board game" in which markers move and collide under a set
of
rules derived from this extension to Lattice Gas theory - mass, momentum, and
energy are conserved exactly and for all time.  Binary-encoded fluid cells, or
"voxels," take the place of floating point numbers as the fundamental unit of
representation, and a small set of primitive operations on voxels takes the
place of ``multiply" and ``add."  This new representation requires 1000 times
fewer bit operations than existing floating-point based methods to compute
comparable results.  This fundamental efficiency can be realized through the
construction of a ``fluids co-processor," the complexity of which is nearly
identical to that of a floating point co-processor.  Many such fluids
co-processors can be interconnected to form a simple, scalable parallel system.

We present the basic principles of the technology and demonstrate
its accuracy by comparing simulation results to laboratory observation for
flows exhibiting separation and vortex shedding.

\begin{center}
{ \bf Transport and Diffusion in a Model Fluctuating Medium} \\
{\bf M. Moreau, B. Gaveau, M. Frankowicz *** and A. Perera*} \\

*Laboratoire de Physique Th\'eorique des Liquides \\
Universit\'e P. et M. Curie, Bolte 121, 75252 PARIS (France) \\
**U. F. R. de Math\'emathiques, Univerist\'e P. et M. Curie, PARIS \\
***Faculty of Chemistry, Jagiellonian University, KRAKOW (Poland) \\
\end{center}

Our purpose is to model the motion of a particle in a time dependent medium
where the fluctuations of the medium are spatially uncorrelated but have a
finite correlation time, so that it is needed to keep the past trajectories of
the particle in memory in order to describe its future evolution.  The medium
is
represented by one and two dimensional lattices.  Each node of the lattice
fluctuates between two internal states according to a random telegraph process.
A particle moves on the lattice and obeys to a given stochastic process between
the nodes.  It is diffused by the nodes, the diffusion law of a node depending
on its internal state.  The model interpolates between a random walk with
persistency and percolation problems, according to the values of the relaxation
frequency and of other parameters.  It can be used for the microscopic theory
of
reaction constants in a dense phase, or for the study of diffusion or
reactivity
in a complex medium.

In different cases, the transmission probability of the medium is computed
exactly.  It is shown that the memory effects decrease the transmission
probability of the medium.  Furthermore, stochastic resonances can occur, an
optimal transmission being obtained for a convenient choice of parameters.  In
more general situations, approximate solutions are given in the case of short
and moderate memory of the obstacles and shown to agree with numerical results.
The diffusion in an infinite two-dimensional lattice is studies, by computer
simulations and the memory is shown to affect the distribution of the particles
rather than the diffusion law.

\begin{center}
{ \bf Long Memory Effects in the Stress Correlation Function} \\
{\bf Toyoaki Naitoh} \\

School of Business Administration, Senshu University \\
Higashimita, Tama-ku, Kawasaki, 214, Japan \\
and \\
{\bf Matthieu H. Ernst} \\

Institute for Theoretical Physics, The University of Utrecht \\
Princetonplein 5, P. O. Box 80006, 3508 TA, Utrecht, The Netherlands \\
\end{center}

The stress correlation function (SCF) in a one-dimensional cellular
automata-fluid is calculated by computer simulations up to 3000 time steps.
The
results are compared with he 1-D tails $t^{-1/2}$ and $t^{-2/3}$ of bare (BMC)
and
self-consistent (SCMC) mode coupling theories.  The crossover between both
tails
is estimated to occur after $t_{cross} \approx 35000$ time steps.  For $t <
400$ and systems
with $L \geq 500$ sites there is good agreement with BMC-theory for finite
systems.
For $t > 400$ there are signs of faster-than-$1/\sqrt{t}$-decay in the SCF.
The simulated
data for the ``divergent'' transport coefficient at times $t > 400$ are
analyzed in
terms of a crossover function, constructed from SCMC-theory.  However a
quantitative verification of the SCMC-theory is still out of reach.

\begin{center}
{ \bf An Immiscible Lattice Gas in Three Dimensions} \\
{\bf John F. Olson and Daniel H. Rothman} \\

Department of Earth, Atmospheric and Planetary Sciences \\
Massachusetts Institute of Technology \\
Cambridge, MA  02139 \\
\end{center}

We present a simple scheme for constructing collision rules for surface tension
in the framework of the 4-D FCHC lattice gas.  Our rule, constructed in analogy
with earlier work for surface tension in lattice-Boltzmann models, acts to
maximize the difference between the component of pressure normal to interfaces
with the component of pressure transverse to interfaces.  The model exhibits a
phase separation transition at the critical reduced density, $d \approx 0.1$.
We have
measured surface tension and compared it with a crude theoretical
approximation.
We also discuss the isotropy of the surface tension.

\begin{center}
{ \bf Pattern Formation in Phase Transition} \\
{\bf Yue Hong Qian and Steven A. Orszag} \\

PACM, Fine Hall \\
Princeton  University\\
Princeton, NJ  08544, USA \\
\end{center}

There is an increasing interest in modeling complicated phenomena by using
simple models, lattice models (lattice gas, lattice Boltzmann and lattice LBGK)
are examples among others.  Phase transition has been a complicated and
attractive problem.  Most of the numerical simulations are based on Ising
model, which is one of the simplest models in physics. The introduction
of lattice gas  [1]
for incompressible hydrodynamics provides some possibility of studying phase
transition by using a non-local interaction [2].  Phase separation models have
been intensively studied by Rothman's group at MIT [3].  By using lattice BGK
models [4], we are able to demonstrate the existence of phase transition with
or
without surface tension.  Van der Waals equation is one of the examples.  We
are
interested particularly in the pattern formation of droplets or bubbles:  the
dynamical phase transition.  The theoretical critical point and phase diagram
are
confirmed by numerical simulation.  Different ``pseudo-potential'' which can
lead
to a non-monotone pressure in function of density are used to test the
sensitivity of the behavior near the critical point.  A universal scaling with
an exponent -1/2 is obtained for the first time [5].  Numerical results also
concern the surface tension and correlation function of density.  We present at
the same time models for one, two and three dimensions.  We will  discuss the
applications and generalizations of our models.  The treatment of wettability
will be included and the multi-fluid systems without the optimization procedure
will be outlined.

References

\noindent
1.	U. Frisch, B. Hasslacher and Y. Pomeau, {\em Phys. Rev.  Lett.},  56, 1505
(1986). \\
2.	C. Appert and S. Zaleski, {\em  Phys. Rev. Lett.}, 64, 1 (1990). \\
3.	A. K. Gunstensen and D. H. Rothman, {\em Physica D}, 47, 47 (1991) \\
4.	Y. H. Qian, D. d'Humi\'eres and P. Lallemand, {\em Europhys. Lett.} 17 (16),
479 \\
(1992); H. D. Chen, S. Y. Chen and W. H. Matthaeus, {\em Phys. Rev.} A 45, 5339
\\
(1992); H. D. Chen and X. W. Shan to be published. \\
5.	Y. H. Qian and S. A. Orszag, to be published in {\em J. Scientific
Computing.} \\

\begin{center}
{ \bf Correlation of Experimental Data with Computational
Data Generated by Fine-Grain, Fixed-Grid Calculations}

{\bf Mr. Peter P. F. Radkowski III}

Radkowski Associates \\
P. O. Box 1121 \\
Los Alamos, NM  87544 \\
\end{center}

Fine-grain, fixed-grid calculations have been correlated with recent (2/93)
hypervelocity impact debris clouds.  The non-equilibrium processes of the
debris
cloud formation (for example, the contemporary formation of (i) exothermic
sublimation products and (ii) inert, solid fragments) greatly hinder the
modeling of critical local post-perforation effects (for example, the
discontinuous application of debris cloud impulses).  The author uses the
interactions of a large population of particles to model the initial impact,
chemical reaction, and post-perforation characteristics of the observed test
phenomena.  Correlation with measured test data include:  axial and radial
velocity and momenta; and mass distribution (comparison of actual dynamic
radiographs and simulated computational radiographs).  Traditional Eulerian
calculations are included to highlight the performance of the fine-grain model.
Ongoing and near-term utilization of fine-grain calculations to model other
experimental data will be summarized.

\begin{center}
{\bf Structure of Shock in Boghosian-Levermore Automaton} \\
{\bf K. Ravishankar} \\

Department of Mathematics\\
SUNY\\
New Platz, N.Y \ 12561, USA \\
\end{center}

We classify the stationary measures of the Boghosian- Levermore automaton
and study the hydrodynamics in Euler regime for the asymmetric case.  We
prove a law of large numbers for the location of the microscopic shock position
when the initial profile is a step function.  The results are obtained
using the coupling methods introduced by Ferrari, Kipnis and Saada for the
asymmetric simple exclusion.

\vspace{1.0true cm}

\begin{center}
{ \bf    Self-Organization Induced by a Differential Flow} \\
{\bf       A. Rovinsky and M. Menzinger} \\

Department of Chemistry\\ University of Toronto\\ Toronto, Ont.
	  M5S 1A1, Canada \\
\end{center}

A new mechanism is described that is believed to play an important role
in the generation of spatiotemporal patterns in natural and artificial
systems. A differential bulk flow of key species destabilizes the
homogeneous steady state of certain kinetic systems with feedback and
gives rise to traveling waves. The Differential Flow Induced  Chemical
Instability (DIFICI) is related to the Turing instability which is
recognized for its central role in development and morphogenesis. DIFICI
may occur in a broad class of systems, from chemical to biological and
ecological.

\begin{center}
{\bf   Simulation of Fines Migration and
 Accumulation in Two Dimensional Porous Media Using Cellular Automata} \\
 {\bf  Maurice Shevalier and Ian Hutcheon} \\

   Department of Geology and Geophysics\\
   The University of Calgary		 \\
   Calgary, Alberta \\
   T2N 1N4, Canada  \\

\end{center}
        The migration and accumulation of fines in pore spaces is a major
problem for the oil industry as it can lead to a decrease in the porosity
and permeability of the reservoir, which ultimately causes an increase in
operating costs and possibly lower rates of production. The purpose of this
study is to simulate fines migration and accumulation in two dimensional
porous media using a cellular automaton.

        Experimental work on fines migration in two dimensional glass micro-
models was carried out (Hutcheon et al. 1989, Goldenberg et al. 1989). From
this work it was found that fines tend to accumulate in the pore spaces,
causing a decrease in the permeability. Also, it was found that particle
migration is controlled by fluid composition and velocity. Further, when
bubbles
are introduced into the system, particle redistribution and transportation
occurs. Finally, it was found that structures of clay particles can form across
the pore throats along the gas-liquid interfaces resulting in a decrease in the
permeability.

        To date, there has not been a comprehensive fundamental study of fines
migration in porous media. The purpose of this study is to simulate fines
migration and accumulation in a two dimensional porous media using a cellular
automaton.

        This study will consist of the study of fines migration in a fluid.
To date
the study has considered fluid flow in porous media as well as the flow of
"simple"
fines particles, i.e. particles that do not interact electrostatically with
each
other. Future work will study the interaction of the fines with the
walls of the porous media as well as with each other, the interaction between
bubbles and fines particles and the effect solution composition has on both
fines and fines-bubble migration.

        Fundamental principles of fluid dynamics, fluid chemistry, colloid
science and electrostatic attraction will be applied to model fines migration
in a porous media. It hoped that a fundamental understanding of fines transport
in two dimensions will be obtained.

 References\\
Goldenberg, L.C., Hutcheon, I.E., and N. Wardlaw,(1989), Experiments on
Transport
of Hydrophobic Particles and Gas Bubbles in Porous Media, {\em Trans. Porous
Media},
4, pp 129-145.

Hutcheon, I.E., Goldenberg, L.C., Nahnybida, C. and M. Shevalier,(1989), Flow
Imparement in Reservoirs,
AOSTRA Final Report, Contract 576, December 1989, 57p.

\begin{center}
{ \bf   Minimally Constrained Lattice Gas Model} \\
{\bf Doug Shim, Tim Spanos, David McEhlaney and Norman Udey} \\

Department of Physics\\
University of Alberta\\
Edmonton\\
Alberta, Canada T6G 2J1. \\
\end{center}

A lattice gas model has been constructed on a 2-D lattice based
solely on the premise that by removing all constraints except
for conservation of energy, momentum, and mass, the macroscopic
physical description will appear naturally. In this model the
events which occur at lattice sites represent the probability
of an event occurring in a region of space ascribed to that
point and during an interval of time. The model allows for
floating-point speeds, and the particles although constrained
to move along the lattice acquires floating-point trajectories.
At present we have demonstrated  that the particle speeds
evolve to a stationary 2-D ``Maxwell-Boltzmann" type distribution
yielding a clear definition of temperature and the concept of an
internal energy. The flow described by this model is identical to
the predictions of the Navier-Stokes equation. Simulation runs on
this model include phase-separation of two immiscible fluids,
Rayleigh-Taylor instability, flow through 2-D porous medium, and
diffusion-limited aggregation growth.
\vspace{1.0true cm}

\begin{center}
{ \bf Near-Criticality in a Lattice Predator-Prey Model} \\
{\bf B. R. Sutherland and A. E. Jacobs }\\

        Deptment of Physics\\ University of Toronto \\
        Toronto, Ontario M5S 1A7, Canada \\
\end{center}

We study the dynamics of ``Wa-Tor'', a predator-prey model on a square
lattice; the parameters are the breeding rates of the predators and prey,
and the predator starvation rate.
The model is robust, evolving to oscillatory, phase-shifted populations
of both species for a large range of parameter values, although the
frequency spectrum is broad due to the randomness of the model.
The most interesting result of the simulations
is that the distribution function $D(s)$ of the prey cluster sizes $s$
is almost critical at large cluster sizes; it is well described by the
power-law form $D(s)\propto s^{\beta}$, which
is, however, cut off at a size
$s_{\mbox{co}}$ (generally greater than $1000$); the exponent
$\beta\simeq 1.3\pm 0.2$
is weakly dependent on the parameters of the model.
The age distribution of the
prey is exponential, with the
lifetime essentially a function of the prey breeding rate alone.

\newpage
\begin{center}

{\bf Simulation Validations of Flows
 Around Circular Cylinders and the Ahmed Body 	} \\
{\bf Chris Teixeira} \\

Exa Corporation\\
125 Cambridge Park Drive\\
Cambridge, MA. 02140, USA  \\
\end{center}

Simulation studies using a lattice gas algorithm (LGA) are presented for
circular
cylinder flows over a wide range of Reynolds numbers (Re). The LGA used has the
property that it effectively erases the underlying lattice from the
macrodynamics
allowing the model to reproduce the results of continuum hydrodynamics
exactly.  For low Re flows ($Re < 100$), we demonstrate the accurate
reproduction
(to within the error of experimental observation)
of drag coefficients and eddy-length to diameter ratios for$ Re < 45$ and
accurate Strouhal number reproduction for $Re > 45$ where vortex shedding
occurs.
The onset of vortex shedding from a steady system occurs naturally for this
LGA at
a $Re = 45$, the same value as found experimentally.
This is in contrast to CFD results which show significant discrepancies with
experimental results in this range of Re and require artificial perturbation of
the flow in order to initiate shedding.  Accurate reproduction of flow
properties
around a cylinder and an Ahmed body (crude car shape) at Reynolds numbers
of practical interest, ($Re ~ 10^6$) will also be presented.

\vspace{1.0true cm}

\begin{center}
{ \bf Monte Carlo Techniques in the Lattice-Boltzmann Applications} \\
{\bf Adnan Taymaz$^1$ and Ayse Zehra Aroguz$^2$} \\

$^1$Physics Department, Faculty of Science Vezneciler Campus \\
$^2$Chemistry Department, Faculty of Engineering, Avcilar Campus \\
Istanbul University 34459 Istanbul, Turkey \\
\end{center}

The energy fluence parameters of interest at given points in a medium have been
calculated using Monte Carlo simulation.  The Boltzmann equation solution for
given physical phenomenon has been obtained.  The uncertainty in Boltzmann
equation solution of a physical model have also been analyzed and compared in
two dimension with the Boltzmann transport equation solution.  In the
calculation a series of sampling techniques is used and better sampling
techniques has been developed.

\begin{center}
{ \bf Application of the Lattice Boltzmann Gases to Cooling Down of Cut
Flowers}
\\
{\bf R. van der Sman} \\

Agrotechnological Research Institute \\
Wageningen, The Netherlands \\
\end{center}

For modeling the cooling down process of packaged cut flowers by forced air
convection the technique of Lattice Boltzmann Gasses is investigated.  The
technique seems to suit our requirements for the modeling technique.  These
requirements are: \\
\noindent
$\bullet$ The modeled system can be built with elementary building blocks with
local
interaction. \\
$\bullet$ Easy refinement of the model is possible. \\
$\bullet$ The modeling technique is generable applicable to transport
phenomena. \\
$\bullet$ Complicated geometries of packages are easy to model. \\
$\bullet$ The modeling technique should have an intuitive feel; {\bf i. e.}
should have a
resemblance of the conceptual model of the scientist. \\

A 1-dimensional and 2-dimensional model describing the velocity field and the
heat and water vapor transport will be presented.  The cut flowers will be
modeled as a porous medium, for which Darcy's law is applied.  Transport of
heat
and water vapor from the flowers will be modeled as scalars convected by the
velocity field.

\begin{center}
{ \bf Commutation of Cellular Automata Rules} \\
{\bf Burton Voorhees} \\

Faculty of Science \\
Athabasca University \\
Box 10,000 \\
Athabasca, AB \\
CANADA T0G 2R0 \\
\end{center}

Let X be the global operator representing a given cellular automata rule.  It
is
shown that the set of all cellular automata rules which commute with X is
determined by the solution set of a system of non-linear Diophantine equations.
Some consequences of this are discussed.

\begin{center}
{\bf Pattern Formation in Lorentz Lattice Gas Cellular Automata} \\
{\bf F. Wang, E. G. D. Cohen} \\

The Rockefeller University \\
1230 York Avenue, \\
New York, NY  10021, USA \\
\end{center}

Previous investigations of Lorentz Lattice Gas Cellular Automata (LLGCA)
involved the motion of a point particle along the bonds of a lattice, whose
sites were randomly occupied (fully or partly) by two types of scatterers:
either left or right reflecting mirrors or left or right turning rotators.  New
types of diffusive behavior of the particles through the scatterers were
discovered[1,2,3].  In case the lattice is fully occupied by only one type of
scatters, the motion of the particle shows pattern formation, where propagation
or other types of regular motion occur.  A number of examples will be given.

References\\
\noindent
1.	X. P. Kong, E. G. D. Cohen, {\em Phys. Rev.} B, 40, 4838 (1989). \\
2.	X. P. Kong, E. G. D. Cohen, {\em J. Stat. Phys.}, 62, 737 (1991). \\
3.	F. Wang, E. G. D. Cohen, ``Diffusion in Lorentz Lattice Gas Cellular
Automata:
the hexagonal and quasi-lattices compared with he squares and triangular
lattices", (to be published).

\newpage

\begin{center}
{\bf New Class of Cellular Automata for
Reaction-Diffusion Systems Applied to the CIMA Reaction} \\
{\bf Joerg Richard Weimar} \\

Universit\'{e} Libre de Bruxelles		\\
   Service du Chimie Physique, C.P. 231\\
        Universit\'{e} Libre de Bruxelles\\
                B-1050 Bruxelles, Belgium
\end{center}

I present a class of cellular automata (CAs) for modeling
reaction-diffusion systems. The construction of the CA is general
enough to be applicable to a large class of reaction-diffusion
equations. The automata are based on a running average procedure to
implement diffusion, and on a probabilistic table-lookup to implement
the reaction. As an example application I present the Lengyl-Epstein
model for the chlorite-iodide-malonic acid reaction (CIMA), which
exhibits a rich set of behaviors: oscillations, hexagonal structures,
stripes, oscillations, and spirals. I investigate cases showing mixed
states, in which different structures coexist in space: isolated
spots, isolated regions of hexagons in a surrounding homogeneous
region, coexistence between stripes and oscillations, and hexagons
and stripes.  The cellular automaton approach has the following
advantages: fast simulations of large systems, simple introduction of
noise in the system, and the possibility find the connections to
other, more phenomenologically constructed CAs.

\begin{center}
{\bf A Single Lattice-Gas Cellular Automaton Model for Lasers} \\
{\bf Xiao-Guang Wu} \\

Department of Chemistry\\ University of Toronto\\ Toronto, Ontario,
	  Canada M5S 1A1 \\
\end{center}

 A lattice-gas cellular automaton model is constructed for
perfectly tuned, one-mode lasers. The model is fully discrete.
A photon representation is used to describe the electromagnetic field.
The atom-field interaction is treated by probabilistic updating rules
that are designed on the basis of rate equation theory. This model
incorporates diffusive motion (random walk with exclusion) of active
particles in the dynamics so one is able to study diffusion effects on
spatial hole burning.
\vspace{1.0true cm}

\begin{center}
{\bf Thermohydrodynamic Lattice-Gas Simulation on the CAM-8} \\
{\bf Jeffrey Yepez$^{*}$ } \\

 Atmospheric Sciences Division, Phillips Laboratory\\
                 Hanscom AFB, MA 01731-5000, USA \\
\end{center}

Much progress has been made in recent years towards developing
lattice-gas automata (LGA) and lattice Boltzmann methods for modeling
hydrodynamic systems.  Multiphase models have shown the ability of
lattice-gases to undergo a liquid-gas phase transition where the
lattice-gas simulates an attractive central force giving rise to a van
der Waals type equation of state [1].  Multispeed models have also
been shown to produce thermohydrodynamic behavior--the report of such
a multispeed lattice-gas showing a Rayleigh-B\'enard convective
instability is a good example [2]. Very recently, the MIT cellular
automata machine (CAM-8) prototype has been constructed and offers an
economical computational opportunity where lattice-gas simulations can
be run at high site-update rates, on large spaces, with video-rate
display graphics.  These several confluent events have stimulated us
to investigate the practicality of the lattice-gas methodology for
simulating certain ``messy'' aspects of atmospheric dynamics.  Our
initiative explores lattice-gas methods related to atmospheric
dynamics and, will involve the construction of a billion-site
massively parallel CAM.  Here we present a two-dimensional test case
where we have implemented on the prototype CAM-8 a multispeed LGA
model with gravitational forcing, temperature sources and sinks, and
free-slip and no-slip boundaries.  We illustrate the flow dynamics
with the Rayleigh-B\'enard convective instability.  We present
kinematic shear viscosity measurements in Poiseuille flow for this
thermohydrodynamic gas.  We also present exponential number density
distributions under non-Boussinesq equilibrium conditions.  Finally we
present an implementation of a multiphase thermohydrodynamic gas.

\begin{enumerate}

\item  C. Appert and S. Zaleski, ``Lattice Gas with a
Liquid-Gas Transition", {\em
Phys. Rev. Letts.}, {\bf 64}, No. 1, 1-4 (1990)

\item  S. Chen, H. Chen, G. D. Doolen, S. Gutman, and
M. Lee, ``A Lattice
Gas Model for Thermohydrodynamics", {\em J. Stat. Phys.}, {\bf 62}, Nos.
5/6, 1121-1151 (1991)

\end{enumerate}

$^{*}$ Supported by the Mathematical and Computer Sciences Directorate
of the Air Force Office
of Scientific Research and ARO grant DAAL03-89-C-0038.

\begin{center}
{\bf Simulation of Gas Flow Using a Multi-Speed LBE Model} \\
{\bf Chen Yu      } \\

Thermal Engineering Lab\\
Department of Nuclear Engineering\\
Faculty of Engineering  \\
Phone(Office): 03-3812-2111 \\
The University of Tokyo \\
Tokyo, Japan
\end{center}

  A class of multi-speed lattice Boltzmann models were extended into
the compressible limit. We added one degree of freedom into the sound
speed of the modeled fluid, with the use of rest particles and
particle reservoir. Furthermore, the sound speed is formalized to be
dependent on thermodynamic variable(density) in the way the perfect
gas does in adiabatic case. Hence realistic gas flows under isentropic
condition could be simulated and some unphysical phenomena occurred in
the simulation of compressible flow with the previous models could be
corrected.

\newpage
\begin{center}
\hglue-1.5true cm{\Large\bf         PROGRAM}
\end{center}

 {\bf TUESDAY, JUNE 08, 1993:      }\\

{\bf MORNING SESSION}\qquad Chairperson: B. Alder

\begin{tabbing}
bbbbbbbbbbbb    \=     \kill
8:50--9:00      \>    Opening Remarks:
         Director of The Fields Institute\\
	        \>\\
9:00--10:00     \>  M. Henon, CNRS Observatoire de Nice, France		 \\
                \> {\bf ``Lattice Gas without
Semi-Detailed Balance''}	   \\
	        \>\\
 10:00--10:20   \>	COFFEE BREAK  \\
	        \>\\
 10:20--11:20   \>  F. Hayot, Ohio State University, USA     \\
                \> {\bf  ``Vortex Street in L\a'{e}vy Walk''}	 \\
		\>\\
 11:20--12:20   \>  A. Noullez, Princeton University, USA   \\
                \> {\bf ``Collision Rules for Two-Dimensional
Hydrodynamics''}\\
		\>\\
 12:20--1:40    \>   LUNCH BREAK	\\
		\>\\
  1:40--3:10    \> {\bf POSTER SESSION: THE FIELDS INSTITUTE}\\
\end{tabbing}

 {\bf AFTERNOON SESSION}\qquad Chairperson: P. Lavallee

\begin{tabbing}
bbbbbbbbbbbb    \=     \kill
3:20--4:20      \>  N. Margolus, Massachusetts
Institute of Technology, USA \\
                \>         {\bf ``CAM-8: A Computer
Architecture Based on Cellular Automata''}\\
		\>\\
4:20--4:40      \>  COFFEE BREAK    \\
		\>\\
4:40--5:40      \>  S. Succi, IBM Center for Scientific and Engineering
                        Computing, Italy 			\\
                \> {\bf   ``Two and Three-Dimensional Simulations of
         Rayleigh-B\a'{e}nard}\\
                \> {\bf Turbulence with Lattice Boltzmann Method''} \\
		\>\\
5:40--6:40      \> G.D. Doolen, Los Alamos National Laboratory, USA  \\
                \> {\bf  ``Future Computers and Lattice Methods;}\\
                \>  {\bf Multiphase Flows Through Porous Media''}	   \\
		\>\\
  7:00--        \>      {\bf DINNER AT THE WATERLOO INN}\\
\end{tabbing}

  {\bf WEDNESDAY, JUNE 09, 1993:     }	       \\

{\bf  MORNING SESSION}\qquad Chairperson: S. Succi

\begin{tabbing}
bbbbbbbbbbbb    \=     \kill
9:00--10:00     \> D. Rothman, Ecole Normale Sup\a'{e}rieure, France    \\
	        \>  {\bf   ``Surface Tension and
Immiscible Lattice Gases''}\\
	        \>\\
10:00--10:20    \>	COFFEE BREAK			   \\
	        \>\\
 10:20--11:20   \> C. Appert, Universit\a'{e} Pierre et Marie
 Curie, France\\
	        \> {\bf ``Large
Liquid-Gas Models on 2D and 3D Lattices'' } \\
	        \>\\
11:20--12:20    \>  D. d'Humi\a`{e}res, Ecole Normale Sup\a'{e}rieure,
France   \\
 	        \> {\bf ``Exact Solution for Lattice Boltzmann Equation:}\\
                \> {\bf Boundaries
                and Interfaces'' }				       \\
	        \>\\
12:20--1:40     \>   LUNCH BREAK			     \\
	        \>\\
1:40--3:10      \> {\bf POSTER SESSION:\quad THE FIELDS INSTITUTE}\\
\end{tabbing}

{\bf AFTERNOON SESSION}\qquad Chairperson: A. De Masi

\begin{tabbing}
bbbbbbbbbbbb    \=     \kill
3:20--4:20      \>  E.G. Flekkoy, University of Oslo, Norway\\
	        \> {\bf   ``Lattice BGK Model for Miscible Fluid Flow:} \\
                \> {\bf Experiments and Simulations''} \\
		\>\\
4:20--4:40      \>   COFFEE BREAK				\\
	        \>\\
4:40--5:40      \> M. Ernst, University of Utrecht, Netherlands \\
	        \> {\bf   ``Metastability and Pattern Formation} \\
                \> {\bf in Biased
                Lattice Gases''}			\\
		\>\\
 5:40--6:40     \>   R. Rechtman, Universidad Nacional
Autonoma de Mexico, Mexico\\
	        \> {\bf ``Lattice Gas Transport in Porous Media'' }  \\
\end{tabbing}

 {\bf THURSDAY, JUNE 10, 1993:  }				  \\

  {\bf MORNING SESSION}\qquad Chairperson:  R. Desai\\
\begin{tabbing}
bbbbbbbbbbbb    \=     \kill
 9:00--10:00    \>  S.R.S. Varadhan, Courant Institute, USA		   \\
	        \> {\bf ``Asymmetric Mean Zero Random Walk with Exclusion:} \\
                \> {\bf  Self Diffusion'' }  		     \\
		\>  \\
10:00--10:20    \> COFFEE BREAK	     \\
\end{tabbing}

\begin{tabbing}
bbbbbbbbbbbb    \=     \kill
 10:20--11:20   \>  E.G.D. Cohen, Rockefeller University, USA \\
	        \> {\bf	``Propagation and Pattern Formation in Lorentz
                Lattice Gases'' }					 \\
		\>\\
11:20--12:20    \>   A. De Masi, Universit\a`{a} degli
Studi di L'Aquila, Italy\\
	        \> {\bf  ``Glauber Evolution with Kac Potentials'' } \\
		\>\\
 12:20--1:40    \>   LUNCH BREAK 			\\
		\>\\
 1:40--11:00    \>  {\bf TRIP TO NIAGARA FALLS}\\
\end{tabbing}
 {\bf FRIDAY, JUNE 11, 1993:    }\\

 {\bf MORNING SESSION}\qquad Chairperson: M. Moreau\\
\begin{tabbing}
bbbbbbbbbbbb    \=     \kill
  9:00--10:00   \>   X.-G. Wu, University of Toronto, Canada\\
	        \> {\bf  ``Reactive
Lattice-Gas Automata and Chemical Chaos''}\\
		\>\\
 10:00--10:20   \>  	COFFEE BREAK			\\
		\>\\
 10:20--11:20   \> S. Ponce--Dawson, Los Alamos National Laboratory, USA \\
                \> {\bf ``Lattice Boltzmann
Simulations of Pattern Formation}\\
                \> {\bf Reaction-Diffusion Systems''}\\
		\>\\
11:20--12:20    \>  B. Hasslacher, Los Alamos National Laboratory, USA \\
                \> {\bf ``Lattice Gases and New Emergent Complexity} \\
                \>  {\bf in Biochemical Systems'' }     \\
		\>\\
 12:20--1:40    \>   LUNCH BREAK			  \\
\end{tabbing}

 {\bf AFTERNOON SESSION}\qquad  Chairperson: M. Ancona

\begin{tabbing}
bbbbbbbbbbbb    \=     \kill
 1:40--2:40     \>   S. Chen, Los Alamos National Laboratory, USA	  \\
	        \> {\bf ``Lattice Boltzmann Method and its Application} \\
                \> {\bf to Computational Biology''}				 \\
  		\>\\
 2:40--3:40     \>   D. Dab, Universit\a'{e} Libre Bruxelles, Belgium\\
	        \> {\bf  ``Polimerization through Heterogeneous Catalysis:}\\
                \>  {\bf a Reactive Lattice-Gas Automaton Approach''}\\
		\>\\
 3:40--4:00     \>   COFFEE BREAK  \\
		\>\\
 4:00--5:00     \>   R. Monaco, University of Genova, Italy       \\
	        \> {\bf ``The Discrete Boltzmann Equation for Gases with} \\
                \> {\bf  Bimolecular
    or Dissociation-Recombination Reactions''}\\
		\>\\
 5:00--6:00     \> {\bf VIDEO SESSION}\\
\end{tabbing}

\begin{tabbing}
bbbbbbbbbbbb    \=     \kill
 6:10--8:00     \> {\bf DINNER: UNIVERSITY CLUB, UNIVERSITY OF WATERLOO}\\
               	\>\\
8:10--10:00     \>  {\bf ROUND TABLE DISCUSSION:
THE FIELDS INSTITUTE}  \\
\end{tabbing}\vspace{-1.5ex}
             {\bf  Moderator:} B. Alder			   \\
            {\bf   Round Table Members:} B. Boghosian, J.P. Boon,
                       G.D. Doolen, \\
\hglue 48true mm K. Molvig, D. Rothman

\hangindent=15true mm\hangafter=1
 {\bf  Topics:}
               1. Lattice gas automata and statistical mechanics.\\
               2. Lattice Boltzmann vs. lattice-gas models:\\
\hglue 5true mm  their applications to science and engineering.   \\
               3. Lattice gas automata and special-purpose
                  computation.					     \\
               4. Commercial applications of lattice gas automata.  \\

  {\bf SATURDAY, JUNE 12, 1993:  }						 \\

  {\bf MORNING SESSION}\qquad Chairperson: D. d'Humieres
\begin{tabbing}
bbbbbbbbbbbb    \=     \kill
 9:00--10:00    \> E. Presutti,
 II Universit\a`{a} degli Studi di Roma, Italy\\
                \> {\bf    ``Critical Fluctuations in Spin Systems''  }\\
		\>\\
10:00--10:20    \>	COFFEE BREAK			    \\
		\>\\
10:20--11:20    \> J.P. Boon, Universit\a'{e} Libre Bruxelles, Belgium\\
                \>	{\bf	``Fluctuation Correlations
in Lattice Gas Automata''}\\
		\>	       \\
11:20--12:20    \>  B. Boghosian, Thinking Machines, USA \\
 	        \> {\bf	``Correlations and Renormalization in Lattice Gas
                  Automata'' } 	\\
\end{tabbing}
 12:20--2:00\qquad\quad  {\bf
 LUNCH: UNIVERSITY CLUB, UNIVERSITY OF WATERLOO}\\

\vfil
\end{document}